\documentclass[a4paper]{jpconf}

\usepackage{graphicx}
\usepackage{graphicx}
\usepackage{graphicx}
\usepackage{bm}
\usepackage{amssymb}
\usepackage{amsmath}
\def\be{\begin{equation}}
\def\ee{\end{equation}}
\def\bed{\begin{displaymath}}
\def\eed{\end{displaymath}}
\def\bea{\begin{eqnarray}}
\def\eea{\end{eqnarray}}
\def\bear{\begin{array}}
\def\eear{\end{array}}
\def\bes{\begin{subequations}}
\def\ees{\end{subequations}}

\newcommand{\MSbar}{\overline{\rm MS}}  
\newcommand{\m}{{\overline m}}

\newcommand{\U}{{\Upsilon}}
\newcommand{\RS}{\rm RS}
\newcommand{\nn}{\nonumber\\}
\usepackage{graphics}
\usepackage{graphicx}
\usepackage{dcolumn}
\usepackage{bm}
\usepackage{epsfig}
\usepackage{graphicx}
\usepackage{multirow}
\usepackage{dcolumn}
\usepackage{graphicx,epsfig}%

\def\lsim{\mathrel{\rlap{\lower4pt\hbox{\hskip1pt$\sim$}}
    \raise1pt\hbox{$<$}}}         
\def\gsim{\mathrel{\rlap{\lower4pt\hbox{\hskip1pt$\sim$}}
    \raise1pt\hbox{$>$}}}         

\begin{document}
\title{Mass of the bottom quark from Upsilon(1S) at NNNLO: an update\footnote{preprint USM-TH-343; presented at ACAT 2016, 18-22 Jan.~2016, UTFSM, Valpara\'{\i}so, Chile}}
\author{C\'esar Ayala$^1$, Gorazd Cveti\v{c}$^2$, Antonio Pineda$^3$}
\address{$^1$Department of Theoretical Physics, IFIC, University of Valencia - CSIC, Valencia, Spain}
\address{$^2$Physics Department, Universidad T\'ecnica Federico Santa Mar\'{\i}a, Valpara\'iso, Chile}
\address{$^3$Physics Department and IFAE, Universitat Aut\`onoma de Barcelona, Barcelona, Spain}

\begin{abstract}
We update our perturbative determination of $\MSbar$ mass $\m_b(\m_b)$, by including the recently obtained four-loop coefficient in the relation between the pole and $\MSbar$ mass. First the renormalon subtracted (RS or RS') mass is determined from the known mass of the $\Upsilon(1S)$ meson, where we use the renormalon residue $N_m$ obtained from the asymptotic behavior of the coefficient of the 3-loop static singlet potential. $\MSbar$ mass is then obtained using the 4-loop renormalon-free relation between the RS (RS') and $\MSbar$ mass. We argue that the effects of the charm quark mass are accounted for by effectively using $N_f=3$ in the mass relations. The extracted value is $\m_b(\m_b) = 4222(40)$ MeV, where the uncertainty is dominated by the renormalization scale dependence.
\end{abstract}

\section{Introduction}
\label{sec:intr}

The ($\MSbar$) mass of the bottom ($b$) quark,
$\m_b \equiv \m_b(\m_b)$, 
is an important quantity in particle physics, free of renormalon ambiguities, 
and appears in many physical observables. Since it is
relatively high, $\sim 4$ GeV, perturbative QCD methods
are suitable for its extraction. The mass of the ground state of the
$b \bar b$ quarkonium, $\Upsilon (1S)$, 
is one of the best quantities for such an extraction,
$M_{\U (1S)}^{\rm (th)} = 2 m_b + E_{\U (1S)} = 9.460 \ {\rm GeV}$,
where $m_b$ is the pole mass of the bottom quark, 
and 
$E_{\U (1S)}$ is the binding energy.
We use the available perturbative expansions of
$2 m_b/\m_b$ and of 
$ E_{\U (1S)}/\m_b$ in powers of QCD coupling
$a(\mu) \equiv \alpha_s(\mu)/\pi$ and thus extract the value of 
$\m_b$. In the extraction, we use the fact that
the leading infrared (IR) renormalon ambiguity of $2m_b$ cancels
out with that of $E_{\U (1S)}$ \cite{Pineda:1998id,Hoang:1998nz,Beneke:1998rk}.

These proceedings are a brief review of our previous work \cite{JHEP2014},
which we update by including in the analysis the recently calculated 
\cite{Smirnovs} four-loop coefficient of the relation between
the pole mass and the $\MSbar$ mass. 
Here we outline: (1)
The correct treatment of charm quark mass effects in the 
perturbation expansion of
$m_b/\m_b$;
(2)
Asymptotic expressions for the coefficients in the perturbation
expansion of the ratio $m_b/\m_b$ and of the static singlet potential $V(r)$, 
and the extraction of the renormalon residue $N_m$;
(3)
The construction of the (modified) renormalon-subtracted mass $m_{b, \RS^{(')}}$ (using $N_m$), and the renormalon-free relation between $m_{b, \RS^{(')}}$ and $\m_b$;
(4)
Renormalon-free perturbation expansion for $M_{\U (1S)}^{\rm (th)}$ in terms of $m_{b, \RS^{(')}}$, and
extraction, from $M_{\U (1S)}^{\rm (th)} = 9.460 \ {\rm GeV}$,
of the values of $m_{b, \RS^{(')}}$ ($\Rightarrow \m_b$).

\section{Charm mass effects in the bottom pole mass}
\label{sec:charm}

The pole mass $m_b$ and the $\MSbar$ mass $\m_b$ are related:
\be
m_b=\m_b\left( 1+ S(N_f) \right)+\delta m^{(+)}_c\ ,
\ee
\be
\label{Smexp}
{\rm where}  \;\;\;\;\;\;\;\;\; S(N_f)=\frac{4}{3} a_{+}(\mu)
{\big [} 1 + r^{(+)}_1(\mu)  a_{+}(\mu) + r^{(+)}_2(\mu) a_{+}^2(\mu) + r^{(+)}_3(\mu) a_{+}^3(\mu) 
+ {\cal O}(a_{+}^4) {\big ]} 
\ee
and the evaluation is usually performed in QCD with $N_f=N_{l}+1 = 4$
active flavors: $r^{(+)}_j(\mu)\equiv r_j(\mu;N_f)$, $a_{+}(\mu)=a(\mu; N_f)$.
The coefficients $R_0=4/3$ and $r_j$ ($j=1,2$) were obtained in 
Refs.~\cite{Tarrach:1980up}, \cite{Gray:1990yh},
\cite{Chetyrkin:1999ys,Melnikov:2000qh}, respectively.
Recently, numerical values of the 4-loop coefficient $r_3$ were obtained \cite{Smirnovs},
and we incorporate them here in the form given in \cite{Kiyo:2015ooa}.

These coefficients have a specific dependence on the 
renormalization scale $\mu$, dictated by $\mu$-independence of  $S(N_f)$
\be
r_1(\mu; N_f) = r_1(N_f) + \beta_0 L_m(\mu) 
\ ,  \; {\rm etc.}
\label{r1mu} 
\ee
where $L_m(\mu) = \ln(\mu^2/\m_b^2)$, and we maintain,
for simplicity, the notation $r_j \equiv r_j(\m_b)$. We will use the notations
$\beta_0 = (1/4)(11 - 2 N_f/3)$ and
$\beta_1 = c_1 \beta_0 = (102 - 38 N_f/3)/16$ for the first two
coefficients of the RGE of $a(\mu)$
 \be
\frac{d a(Q)}{d \ln Q^2} = - \beta_0 a^2(Q)
\left(  1 + c_1  a(Q) + c_2 a^2(Q) + 
c_3 a^3(Q) +  \cdots \right) \ .
\label{RGE}
\ee

Finite-mass charm quark effects are incorporated in 
\be
\delta m_c^{(+)}=
\delta m_{(c,+)}^{(1)}a_{+}^{2}(\m_b)+
\delta m_{(c,+)}^{(2)}a_{+}^{3}(\m_b)+{\cal O}(a_{+}^{4})
\,,
\ee 
which vanishes in the $m_c \rightarrow 0$ limit. We have
\be
\delta m_{(c,+)}^{(1)}=\frac{4}{3}\m_b\Delta[\m_c/\m_b]= 1.9058\;{\rm MeV}
\; \cite{Gray:1990yh},
\quad \delta m_{(c,+)}^{(2)}=48.6793\;{\rm MeV} \; \cite{Bekavac:2007tk},
\ee
\be
\Rightarrow \;\;\;\;  \delta m_{(c,+)}^{(1)} a_{+}^2(\m_b) = 9.3 \; {\rm MeV},
\qquad
\delta m_{(c,+)}^{(2)} a_{+}^3(\m_b) = 18.1\;{\rm MeV},
\label{dmcdiv}
\ee
so $\delta m_c^{(+)}$ is badly divergent. Why? 
At loop order $n$, the natural scale of the loop integral for $m_b$
is $m_b e^{-n}$ \cite{Ball:1995ni}, which for $n$ large 
enough is: $m_b e^{-n} < m_c$. Therefore, for large $n$ ($>2$) charm quark appears
as very heavy (decoupled), leading to the effective number of
flavors being $N_l =3$ and not $N_f=N_l+1=4$.
Therefore, it is convenient to rewrite the relation between the pole 
and the $\MSbar$ mass in terms of  $a_{-}(\mu)=a(\mu;N_l)$ and 
$r^{(-)}_j(\mu)\equiv r_j(\mu;N_l)$ [$N_l=3$]
\be
m_b=\m_b\left( 1+ S(N_l)\right)+\delta m_c\ ,
\label{SmNl0}
\ee
\bea
\label{SmNl}
{\rm where} \;\;\;\; S(N_l) &=& \frac{4}{3} a_{-}(\mu)
{\big [} 
1 + r^{(-)}_1(\mu)  a_{-}(\mu) + r^{(-)}_2(\mu) a_{-}^2(\mu) + r^{(-)}_3(\mu) a_{-}^3(\mu) 
+ {\cal O}(a_{-}^4) {\big ]},
\eea
and $r_j^{(-)}(\m_b)=7.74$, $87.2$, $1265.3 \pm 16.1$, for $j=1,2,3$.
The effects of the decoupling of $S$ ($N_f \mapsto N_l=3$) are absorbed in the new $\delta m_c$
\bea
\delta m_c &=& \left[\delta m_{(c,+)}^{(1)}+\delta m_{(c,\rm dec.)}^{(1)} \right]
a_{-}^{2}(\m_b)+
\left[\delta m_{(c,+)}^{(2)}+\delta m_{(c,dec.)}^{(2)} \right]
a_{-}^{3}(\m_b)
+{\cal O}(a_{-}^4)
\ ,
\label{delmc}
\eea
where $\delta m_{(c,\rm dec.)}^{(j)}$ are generated by this decoupling and read
\be
 \delta m_{(c,\rm dec.)}^{(1)}=
\frac{2}{9}{\m_b}\left({\rm ln}\left(\frac{\m_b^2}{\m_c^2}\right)-\frac{71}{32}-\frac{\pi^2}{4} \right)
\ee
and $\delta  m_{(c,\rm dec.)}^{(2)}$ can be found in Ref.~\cite{JHEP2014}.

Numerical evaluation gives for 
$\left[\delta  m_{(c,+)}^{(1)}+\delta  m_{(c,\rm dec.)}^{(1)} \right]a_{-}^{2}(\m_b)
= -1.6$ MeV and 
$\left[\delta  m_{(c,+)}^{(2)}+\delta  m_{(c,\rm dec.)}^{(2)} \right]a_{-}^{3}(\m_b)
= -0.3$ MeV.  
This means that the previous divergent series (in ${\rm QCD}_{N_f=4}$)
$\delta m_c^{(+)} = (9.3 + 18.1 + \ldots)$ MeV [Eq.~(\ref{dmcdiv})]
now tranforms (in ${\rm QCD}_{N_l=3}$) to
\be
\delta m_c = (-1.6 - 0.3 + \ldots) \ {\rm MeV}.
\label{dmcconv}
\ee
The series for $\delta m_c$
in ${\rm QCD}_{N_l=3}$ formulation is convergent, 
strong cancellation takes place between $\delta m_{(c,+)}^{(j)}$ 
and $\delta m_{(c,\rm dec.)}^{(j)}$, as expected. 

\section{Leading renormalon of the pole mass}
\label{sec:IRren}

The asymptotic behaviour of $r_N$ is determined by the
leading IR renormalon:
\be
\label{rNasym}
 \frac{4}{3} r^{\rm asym}_N(\mu)  \simeq    \pi N_m \frac{\mu}{\m_b}
(2 \beta_0)^N 
\frac{ \Gamma ( \nu + N + 1) }{\Gamma(\nu + 1) }
\left[ 1 + \sum_{s=1}^3 \frac{ \nu \cdots (\nu - s + 1)}{(N+\nu)\cdots(N+\nu-s+1)} {\widetilde c}_s +
{\cal O} \left( N^{-4} \right) \right].
\ee
\be
\label{hms}
 \frac{4}{3} r_N(\mu)  =    \pi N_m \frac{\mu}{\m_b}
(2 \beta_0)^N \frac{ \Gamma ( \nu + N + 1) }{\Gamma(\nu + 1) }
\left[ 1 + \sum_{s \geq 0} \frac{ \nu \cdots (\nu - s + 1)}{(N+\nu)\cdots(N+\nu-s+1)} {\widetilde c}_s \right] +h_N(\mu),
\ee
where $h_N$ is dominated by subleading renormalons, and the coefficients ${\widetilde c}_s$ ($s=1,2,3$) are given in \cite{Beneke:1994rs,Pineda:2001zq,Contreras:2003zb,JHEP2014} (${\widetilde c}_0=1$ by convention).

Determining the pole mass from $\U (1S)$ mass 
has large uncertainties due to the pole mass renormalon
ambiguity $\delta m_b \sim {\Lambda}_{\rm QCD} $
\cite{Beneke:1994rs}.
In order to avoid this problem, we work with the renormalon-subtracted
(RS) bottom mass $m_{b, \RS}$ instead 
\cite{Pineda:2001zq}.
Then, $\m_b$ is obtained from its stable (renormalon-free) relation 
with the $m_{\RS}$ mass. 

The use of $m_{\RS}$ in the theoretical evaluation of the $\U (1S)$ mass
is convenient  because it has no leading IR
renormalon ambiguity, and the renormalon cancellation in the 
quarkonium mass $M_{\Upsilon(1S)} = 2 m_b + E_{\U (1S)}$ 
is implemented automatically and explicitly.

\section{Determination of the renormalon residue $N_m$ and $N_V$}
\label{sec:Nm}
The asymptotic behavior of the coefficients $v_N(\mu)$ of the static singlet potential,
\be
V(r) =  -\frac{4\pi}{3}\frac{1}{r} a_{-}(\mu) \left[
1\!+\! v_1(\mu) a_{-}(\mu)\!+\! v_2 a_{-}(\mu)^2\! +\! v_3 a_{-}(\mu)^3\!+\!\ldots \right]
\,,
\ee
can be determined in complete analogy with
those of $r_N$
\bea
- \frac{4}{3} v_N(\mu)  &=&    N_V \mu r
(2 \beta_0)^N \sum_{s \geq 0} {\widetilde c}_s
\frac{ \Gamma ( \nu + N + 1-s) }{\Gamma(\nu + 1-s) } 
+d_N(\mu)
\quad \Rightarrow
\label{vN}
\\
\label{vNasym}
  -\frac{4}{3} v^{\rm asym}_N(\mu)  & \approx &  N_V \mu r
(2 \beta_0)^N 
\frac{ \Gamma ( \nu + N + 1) }{\Gamma(\nu + 1) } 
\left[ 
 1 + \sum_{s=1}^3 \frac{ \nu \cdots (\nu - s + 1)}{(N+\nu)\cdots(N+\nu-s+1)} {\widetilde c}_s \right],
\eea
where in Eq.~(\ref{vNasym}) $d_N=0$ was taken. We can determine the ``strength'', $N_V$, of the leading IR renormalon by approximating the asymptotic $v^{\rm asym}_N(\mu)$
with the exact $v_N(\mu)$ ($N=0,1,2,3$): 
$v^{\rm asym}_N(\mu) \approx v_N(\mu) \; \Rightarrow$ 
\be
N_V  \approx  -\frac{4}{3}v_N(\mu){\Bigg /}
{\Bigg \{} \mu r
(2 \beta_0)^N 
\frac{ \Gamma ( \nu + N + 1) }{\Gamma(\nu + 1) }
\left[ 
 1 + \sum_{s=1}^3 \frac{ \nu \cdots (\nu - s + 1)}{(N+\nu)\cdots(N+\nu-s+1)} {\widetilde c}_s \right] {\Bigg \}}.
\ee
The result for
$N_V$ should be the best for the highest available $N$ ($N=3$)
and should also have reduced spurious $\mu$-dependence. At present, the $v_j$ are known up to ${\rm N}^3{\rm LO}$ ($v_3$)~\cite{Fischler:1977yf,Billoire:1979ih,Schroder:1998vy,Pineda:1997hz,BPSV,Kniehl:2002br,Penin:2002zv,Smirnov:2008pn,Anzai:2009tm,Smirnov:2009fh}.

In the sum $2 m_b + V(r)$ the leading IR renormalon gets cancelled. $N_V$ is then related with $N_m$ by
the renormalon cancellation of the sum $2 m_b + V(r)$:
$2 N_m + N_V=0$. Determining $N_m$ via $N_V$ gives us the value that we use
\cite{JHEP2014}
\be
N_m = -  N_V/2 = 0.56255(260)  \; (N_l=3).
\ee

\section{Renormalon-subtracted (RS, RS') mass of bottom}
\label{sec:mRS}

The RS mass is defined by subtracting the leading IR renormalon singularity  from the pole mass \cite{Pineda:2001zq}:
\be
\label{mrs1}
m_{b, \RS}(\nu_f) = m_b - 
N_m \pi \nu_f \sum_{N=0}^{\infty} a_{-}^{N+1}(\nu_f)(2\beta_0)^N 
 \sum_{s \geq 0} {\widetilde c}_s
\frac{ \Gamma ( \nu + N + 1-s) }{\Gamma(\nu + 1-s) }
\ee
Equation (\ref{mrs1}) is still formal. In practice, one rewrites $m$ in terms of $\m$ using Eqs. (\ref{SmNl0})-(\ref{SmNl})
\be
m_b = \m_b (1 + (4/3) a_{-}(\nu_f) + \ldots),
\ee
and reexpands the perturbation series in Eq. (\ref{mrs1}) around the same 
coupling $a_{-}(\mu)$, 
at fixed but otherwise arbitrary scale $\mu$:
\be
m_{b,\RS}(\nu_f) =
\m_b \left[ 1 +\sum_{N=0}^{\infty} h_N(\nu_f) a_{-}^{N+1}(\nu_f) \right]  
\Rightarrow
m_{b,\RS}(\nu_f)=\m_b \left[ 1 +\sum_{N=0}^{\infty} {\widetilde h}_N(\nu_f;\mu) a_{-}^{N+1}(\mu) \right] \ ,
\label{mrs2b}
\ee
where $h_N(\nu_f)$ is determined from Eq.~(\ref{hms})
(with $\mu=\nu_f$ and with the sum truncated at ${\widetilde c}_3$) 
for $N=0,1,2,3$. For $N \geq 4$ we take $h_N(\m_b)=0$.
The coefficients 
${\widetilde h}_N(\nu_f;\mu)$
in Eq.~(\ref{mrs2b}) are obtained by expanding $a_{-}(\nu_f)$ in
the expansion in powers of $a_{-}(\mu)$. 
Note that $m_{b,\RS}(\nu_f)$ will only marginally depend on $\mu$ when we truncate the infinite sum in Eq.~(\ref{mrs2b}). On the other hand, the coefficients $h_N$ are functions of $\nu_f$, $\mu$, and $\m_b$, and are much smaller than $r_N(\mu)$.


A variant of the RS mass is the modified renormalon-subtracted (RS') mass
$m_{b, \RS'}$, where subtractions start at $\sim a^2$ 
\cite{Pineda:2001zq}.
Specifically in this case, Eqs.~(\ref{mrs1}) and (\ref{mrs2b}) are repeated, with the
replacements $m_{b,\RS}(\nu_f) \mapsto m_{b, \RS'}(\nu_f)$ and $\sum_{N=0}^{\infty}
\mapsto \sum_{N=1}^{\infty}$.

\section{Bottom mass from heavy quarkonium}
\label{sec:mb}

The perturbation expansion of $M^{(th)}_{\Upsilon(1S)}$
is presently known up to ${\cal O} (m_b a^5)$
\cite{Pineda:1997hz,BPSV,Kniehl:2002br,Penin:2002zv}:
\vspace{-0.2cm}
\bea
M^{(th)}_{\Upsilon(1S)} &=& 2m_b - \frac{4 \pi^2}{9} m_b a_{-}^2(\mu) 
{\Bigg \{} 1 + a_{-}(\mu) \left[ K_{1,0} + K_{1,1} L_p(\mu) \right] 
+ a_{-}^2(\mu) \sum_{j=0}^2 K_{2,j} L_p(\mu)^j
\nn &&
+ a_{-}^3(\mu) 
{\Big [} 
K_{3,0,0} + K_{3,0,1} \ln a_{-}(\mu)
+ \sum_{j=1}^3 K_{3,j} L_p(\mu)^j 
{\Big ]}
+ {\cal O}(a_{-}^4) {\Bigg \}} \ ,
\label{Ebb}
\eea
$\mu$ is the renormalization scale, 
$L_p(\mu) =  \ln ( {\mu}/\mu_b) $ where $\mu_b=(4 \pi/3) m_b 
 a_{-}(\mu)$. $K_{i,j}(N_f)$ and $K_{3,0,j}$ are given, e.g., in
\cite{JHEP2014}.
We then rewrite $m_b$ in terms of $m_{b, \RS}$ 
to implement the leading IR
renormalon cancellation. This gives
\bea
\lefteqn{\!\!\!\!\!\!
\frac{M^{(th)}_{\Upsilon(1S)}}{ m_{b,\RS}(\nu_f)} =  2
+ \left[ 
2 \pi N_m b a {\cal K}_0 - \frac{4 \pi^2}{9} a^2 
\right]
+ \left[
2 \pi N_m b a^2 \left( {\cal K}_1 + z_1 {\cal K}_0 \right)  
- \frac{4 \pi^2}{9} a^3 \left( K_{1,0} + K_{1,1}L_{\RS}
\right) \right]
}
\nn &&
\!\!\!\!  +\!\left[
2 \pi N_m b a^3 \left( {\cal K}_2\!+\!2 z_1 {\cal K}_1\!+\!z_2 {\cal K}_0 \right)
\!-\!\frac{4 \pi^2}{9} \left( a^4 \sum_{j=0}^2 K_{2,j} L_{\RS}^j\!+\!b a^3 \pi N_m {\cal K}_0
\right) \right] + {\cal O}(b a^4\!,\!a^5).
\label{MUs2a}
\eea
The terms ${\cal O}(b a^4, a^5)$ have a similar structure and were written in
\cite{JHEP2014}.
The notations are
\begin{subequations}
\label{ambLNKN}
\be
a  \equiv  a_{-}(\mu) = a(\mu,N_f=3) \ ; 
\quad b \equiv  b(\nu_f) = \nu_f/m_{b,\RS}(\nu_f) \ , 
\quad N_m  =  N_m(N_l=3) \ ,
\label{abNm}
\ee
\be
L_{\RS} \equiv L_{\RS}(\mu) =
\ln \left( \frac{\mu}{ (4 \pi/3) m_{b,\RS}(\nu_f) a_{-}(\mu) } \right),
\; {\cal K}_N  =  (2 \beta_0)^N  \sum_{s=0}^{3} 
{\widetilde c}_s \frac{\Gamma(\nu+N+1-s)}{\Gamma(\nu+1-s)}  \ .
\label{KN}
\ee
\end{subequations}

In the expression (\ref{MUs2a}) for $M_{\Upsilon(1S)}$, the terms of the same order 
$(\nu_f/m_{b,\RS}) a^n$ and $a^{n+1}$ were combined in common brackets $[ \ldots ]$, in order to
account for the renormalon cancellation. 

If using the RS' mass in our approach  instead, the above expressions
are valid without changes, except that $m_{b,\RS} \mapsto m_{b,\RS'}$ and
${\cal K}_0 \mapsto 0$ (and: $h_0(\mu) \mapsto 4/3$).

We note that we take $N_l=3$ active flavours, as the charm quark mass effects
in the binding energy $E_{\Upsilon(1S)}$ are
negligible \cite{Brambilla:2001qk}.

We extract the bottom masses from the condition $M^{(th)}_{\Upsilon(1S)}
= M^{(exp)}_{\Upsilon(1S)}  ( = 9.460 \ {\rm GeV} )$. 
The error estimates are made assuming $\mu=2.5^{+1.5}_{-1.0}$ GeV
[we varied $\mu$ in Eq.~(\ref{MUs2a}) but not in Eq.~(\ref{mrs2b})],
$\nu_f=2\pm 1$ GeV, $\alpha_s(M_z)=0.1184(7)$ (and decoupling at
$\m_b= 4.2$ GeV and at $\m_c=1.27$ GeV), $N_m=0.56255(260)$, 
and $(4/3) r_3(\m_b;N_l)=1687.1 \pm 21.5$ \cite{Kiyo:2015ooa}.

In RS and RS' approaches we extract, in MeV, respectively
\bes
\bea
\label{MRSdet}
m_{b,\RS}(2{\rm GeV})
&=&
4\,437^{-11}_{+43}(\mu)^{-3}_{+5}(\nu_f)^{-2}_{+2}(\alpha_s)^{-41}_{+41}(N_m)^{-0}_{+0}(r_3);
\\
\label{MMSRSdet}
\Rightarrow 
\;\;\;\;   \m_{b}
&=&4\,217^{-10}_{+39}(\mu)^{-3}_{+5}(\nu_f)^{-5}_{+5}(\alpha_s)^{-1}_{+1}(N_m)^{-4}_{+4}(r_3).
\\
\label{MRSprimedet}
m_{b,\RS'}(2\;{\rm GeV})&=&4\,761^{-16}_{+41}(\mu)^{-3}_{+5}(\nu_f)
^{+4}_{-3}(\alpha_s)^{-26}_{+26}(N_m)^{-0}_{+0}(r_3);
\\
\label{MMSRSprimedet}
\Rightarrow 
\;\;\;\;   \m_{b}
&=&4\,223^{-14}_{+36}(\mu)^{-2}_{+4}(\nu_f)^{-4}_{+4}(\alpha_s)^{-1}_{+1}(N_m)^{-4}_{+4}(r_3).
\eea
\ees
The uncertainties in $\m_b$ are dominated by the variation of the 
renormalization scale $\mu$. 

The renormalon cancellations are reflected numerically
in Eq.~(\ref{MUs2a}) [we take $\mu=2.5$ GeV]:
\bes
\bea
 {\rm RS: \;} M_{\Upsilon(1S)}&=& 
 ( 
8874+431+167+18-30
 ) \ {\rm MeV} \,, 
\\
{\rm RS': \;} M_{\Upsilon(1S)}&=&
 ( 
9521-150+112+8-31
 ) \ {\rm MeV} \,, 
\eea
\ees
The convergence is good; except for the last (NNNLO) term
${\cal O}(a^5, b a^4)$, where the factorization scale dependence becomes stronger, which may signal the 
importance of ultrasoft effects.

The relations between RS (RS') mass and $\MSbar$ mass are reasonably convergent:
\bes
\bea
m_{b, \RS}(2\;{\rm GeV})&=&
 ( 
4217+191+36+12-19
 ) \ {\rm MeV} \ ,
\\
m_{b, \RS'}(2\;{\rm GeV})&=&
 ( 
4223+478+60+18-17
 ) \ {\rm MeV} \ , 
\eea
\ees
where the expansion parameter is taken to be $a(2.5\;{\rm GeV})$. A bigger value for the renormalization scale, closer to the bottom quark mass, makes the last term smaller.

Until now we have approximated  $\delta m_c=0$ in Eq.~(\ref{mrs1}). 
However, $\delta m_c \approx -2$ MeV, Eq.(\ref{dmcconv}).
Hence, we have to add $2 \ {\rm MeV}$ to the values of $\bar m_b$ obtained in Eqs.~(\ref{MMSRSdet}) and (\ref{MMSRSdet}) 
(in Ref.~\cite{JHEP2014} it 
was incorrectly subtracted), leading to
the final average of the RS and RS' extractions
\be
\m_b  =  4222(40) \; {\rm MeV} \ .
\label{result}
\ee 
where we have rounded the $\pm$ variation of each parameter to the maximum and added them in quadrature. 
\section{Conclusions}
\label{sec:concl}

\begin{enumerate}
\item
We presented strong numerical indications that the charm quark decouples in the relation between $m_b$ and $\m_b$ ($\Rightarrow N_l=3$). 
\item
An improved determination of the residue of the leading renormalon for the
bottom pole
mass (and static potential with $N_l=3$) was performed: $N_m = 0.56255(260)$.
\item
Use of the 3-loop ($\sim a^5 \m_b$) corection to the $\U (1S)$ binding
energy, and 4-loop relation between $m_b$ and $\m_b$, allowed us to perform extraction of $m_{b, \RS^{(')}}$ and $\m_b$ to
NNNLO, with the resulting values Eq.~(\ref{result}). The uncertainties are dominated by the variation of the renormalization scale.

\end{enumerate}
  
\vspace{0.5cm}

\noindent
{\bf Acknowledgments}

\noindent
This work was supported in part by FONDECYT (Chile) Grant No.~1130599 (GC);
Spanish Government and ERDF funds (EU Commission)
FPA2014-53631-C2-1-P and by CONICYT ``Becas Chile'' Grant No.~74150052 (CA); and 
Spanish Government  FPA2014-55613-P, FPA2013-43425-P (AP).

\end{document}